# A Radio Frequency Non-reciprocal Network Based on Switched Acoustic Delay Lines

Ruochen Lu, *Student Member, IEEE*, Tomás Manzaneque, *Member, IEEE*, Yansong Yang, *Student Member, IEEE*, Liuqing Gao, *Student Member, IEEE*, Anming Gao, *Student Member, IEEE*, and Songbin Gong, *Senior Member, IEEE*

*Abstract*—This work demonstrates the first non-reciprocal network based on switched low-loss acoustic delay lines. The 4-port circulator is built upon a recently reported frequency-independent, programmable, non-reciprocal framework based on switched delay lines. The design space for such a system, including the origins of the insertion loss and harmonic responses, is theoretically investigated, illustrating that the key to better performance and low-cost modulation signal synthesis lies in a large delay. To implement a large delay, we resort to in-house fabricated low-loss, wide-band lithium niobate (LiNbO$_3$) SH0 mode acoustic delay lines employing single-phase unidirectional transducers (SPUDT). The 4-port circulator, consisting of two switch modules and one delay line module, has been modularly designed, assembled, and tested. The design process employs time-domain full circuit simulation and the results match well with measurements. A 18.8 dB non-reciprocal contrast between insertion loss (IL = 6.6 dB) and isolation (25.4 dB) has been achieved over a fractional bandwidth of 8.8% at a center frequency 155 MHz, using a record low switching frequency of 877.19 kHz. The circulator also shows 25.9 dB suppression for the intra-modulated tone and 30 dBm for IIP3. Upon further development, such a system can potentially lead to future wideband, low-loss chip-scale nonreciprocal RF systems with unprecedented programmability.

## I. INTRODUCTION

RECENTLY, full duplex radios have been extensively researched due to their potential in doubling the available bandwidth in wireless communication and radar applications [1]–[3]. The main hurdle for implementing a system capable of simultaneous transmit and receive (STAR) in the same frequency band is the pronounced self-interference from the transmitter (TX) to the receiver (RX) [4]. One of the solutions is to apply a non-reciprocal network, e.g. circulators and isolators, between TX, RX and the antenna (ANT), as seen in Fig. 1. Strong non-reciprocity is preferable in such a network for achieving a high isolation in the TX to RX path while maintaining low insertion losses in the TX to ANT and ANT to RX paths. Conventionally, non-reciprocity is obtained by stripline circulators based on magnetically biased ferrite materials in which the phase velocities of electromagnetic waves traveling in the opposite directions are different [5]–[7]. The transmission and isolation nodes are established by the constructive and destructive interference between the clock wise and counter-clockwise traveling waves. Nevertheless, the dimensions of such a device is set by the wavelength of electromagnetic wave in the magnetic material, which prevents further miniaturization. Another type of ferrite circulators, namely lumped-element based ferrite circulators, can overcome the aforementioned size limits and be one order of magnitude smaller than the wavelength, while simultaneously providing low insertion loss, broad bandwidth, and great linearity [8]–[10]. However, these components are fixed and cannot be reconfigured in the field for dynamic adaptation to the RF ambiance. Moreover, the ferrite circulators cannot be singularly expanded to more than 4 ports without resorting to networking multiple 3 or 4-port circulators. Hence, the inadequate programmability and expandability further limit the extension of these devices for future wireless systems with growing complexity and higher numbers of RF paths (e.g. MIMO).

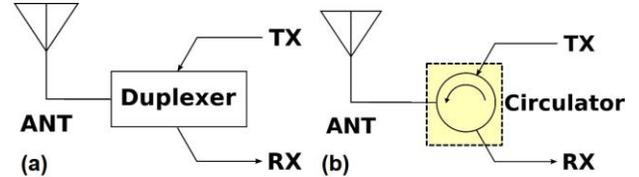

Fig. 1. Notional scheme for (a) a conventional transceiver (composed of switches or filters) and (b) a STAR transceiver using a circulator.

Driven by achieving more integrated and sufficiently linear non-reciprocity in microwave applications, magnet-free non-reciprocal systems based on modulation of reactance [11]–[19] or conductance [20], [21] have recently been explored. In such approaches, the waveguide properties are spatio-temporally modulated to create momentum-biasing similar to the Faraday effect in ferrites for breaking the reciprocity [22]. However, the bandwidth of these applications are limited because they rely on wave interference or mode splitting within a resonant structure, which is a narrow band phenomenon. Furthermore, it is challenging to maintain linearity at high signal levels or

Manuscript submitted March 1$^{st}$, 2018. This work was supported by DARPA MTO NZERO and SPAR programs. Ruochen Lu, Tomás Manzaneque, Yansong Yang, Liuqing Gao, Anming Gao, and Songbin Gong are with the Department of Electrical and Computing Engineering, University of Illinois at

Urbana-Champaign, Urbana, IL 61801 USA (email: rlu10@illinois.edu, tmanzane@illinois.edu, yyang165@illinois.edu, lgao13@illinois.edu, agao4@illinois.edu, and songbing@illinois.edu).

modulation levels due to the intrinsic nonlinearities in the common modulatable elements (e.g. varactors). Another time varying approach for radio frequency (RF) non-reciprocity is to use switched delay lines [23] or sequentially switched delay lines [24], which provides much wider bandwidth performance and potentially excellent linearity. Lately, the authors have developed a new architecture for frequency independent, broadly programmable non-reciprocal network with arbitrary number of ports using switches and delay line arrays [25] with great potentials in multiple input multiple output (MIMO) communication systems. Nevertheless, the network requires the delay in the delay lines to be in the same order of magnitude as the period of the modulation signals. Due to the fast velocities of electromagnetic waves, it requires either high-frequency modulation signals or greatly reduced wave velocity to fit the system into a compact size required by modern wireless devices. The first approach has the drawback of dramatically increased power consumption and system complexity due to the high-frequency modulation signal synthesis and phase matching [26]. The second approach usually adopts slow wave structures in electromagnetic (EM) wave guides [27], [28], however the slow-wave factor is nowhere near adequate for miniature systems.

One way for achieving significant delays in a small form factor is to leverage the orders of magnitude slower velocities of the acoustic waves [29]. However, the high insertion loss (IL) and narrow bandwidth of the traditional surface acoustic wave (SAW) devices prevents meaningful implementations [30], [31]. Recently, acoustic delay lines based on suspended single crystal X-cut lithium niobate (LiNbO₃) thin films have been demonstrated with low-loss wide-band performance [32]–[35] based on the fundamental shear-horizontal (SH0) [36] and length extensional (S0) [37] modes, thanks to the advances in thin film transfer techniques [38]–[40]. The excellent performance is the result of a very high electromechanical coupling ($k_t^2$) [41]–[44] and low attenuation [45]–[48] in the suspended LiNbO₃ thin film. A remarkable delay of 80 ns is reported for a 0.5 mm delay line with a 2 dB IL over a 10% fractional bandwidth (FBW) [49]. Such performance enables the implementation of programmable non-reciprocal networks based on acoustic devices.

This work demonstrates the first non-reciprocal network based on switched acoustic delay lines. A quantitative analysis on the insertion loss, return loss, and harmonic performance due to non-idealities in the system is presented. Based on our analysis on the significance of long delay in such a system, a 4-port circulator system is designed with acoustic delay lines. To validate our design, such a system is implemented with switch modules and delay line modules. The measurement shows a highly symmetric performance across the 4-ports with 18.8 dB non-reciprocal contrast between IL (6.6 dB) and isolation (25.4 dB) over a FBW of 8.8% at a center frequency 155 MHz, using a record low switching frequency of 877.22 kHz. The system also possesses 25.9 dB difference for the modulated tone and 30 dBm for IIP3.

This paper is organized as follows. Section II provides a general discussion on the frequency independent non-reciprocal network with a quantitative analysis of the IL, return loss, isolation, and intra-modulated tones in the system. Section III introduces the designed 4-port circulator based on acoustic delay lines. The design and testing results of the switch module and the acoustic delay line module will be presented. System level simulation is performed afterwards. Section IV offers the measurement results and discussions on the implemented 4-port non-reciprocal network, including S-parameters, group delays, harmonic responses and nonlinearities of the system. Finally, Section V summarizes the conclusion.

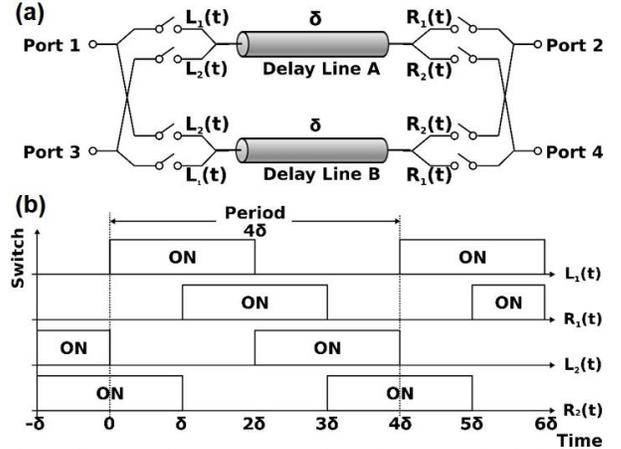

Fig. 2. (a) Concept of a 4-port non-reciprocal network. (b) Switch control waveforms applied to the network for producing the non-reciprocity.

## II. THEORETICAL APPROACH

### A. Frequency Independent Nonreciprocal Network

The general schematic of our frequency independent, broadly programmable non-reciprocal network has been proposed in our previous work [25]. In this work, we focus on a specific implementation, a 4-port circulator based on switched acoustic delay lines, to lower the temporal effort for maintaining the non-reciprocity. The schematic of the 4-port non-reciprocal system [Fig. 2 (a)] consists of two delay lines (A and B) and four single pole single throw switches. The switches are controlled by four control signals [Fig. 2 (b)] that are represented as:

$$L_1(t) = \sum [H(t/4\delta - n) - H(t/4\delta - n - 0.5)] \quad (1)$$
$$R_1(t) = \sum [H(t/4\delta - n - 0.25) - H(t/4\delta - n - 0.75)] \quad (2)$$
$$L_2(t) = \sum [H(t/4\delta - n - 0.5) - H(t/4\delta - n - 1)] \quad (3)$$
$$R_2(t) = \sum [H(t/4\delta - n - 0.75) - H(t/4\delta - n - 1.25)] \quad (4)$$

where $H$ is the Heaviside function with $H(0)$ defined as 1. The switch is on when the control signal equals to 1, and is off when the control signal equals to 0. The control signals have a period ($4\delta$) that is four times the delay line's group delay ($\delta$). Control signals (labeled in Fig. 2) on opposite sides of the delay lines are offset by $\delta$ (or 90° phase difference).

The operation of the circulator can be explained intuitively with bounce diagrams (Fig. 3) [50]. These diagrams demonstrate the transmission of signals along the delay lines (x-axis) in the time domain (y-axis) when all ports are transmitting and receiving simultaneously. The bounce diagrams are used to show how signals would traverse the delay line A and B in the forward (+z) and reverse (-z) directions (Fig. 3) over a period of $6\delta$ with the states of switches $L_I(t)$ and $R_I(t)$ displayed on the corresponding sides. The delay line impedance is assumed to be matched to the port impedance, while the switches are assumed perfectly open in the off-state and perfectly short in the on-state. The switches are also assumed to take 0 time to switch on or off. In operation, the signals fed to Port 1 are first time-multiplexed onto the two delay lines and subsequently demultiplexed to Port 2 by turning on the switches connected to Port 2 at the moment exactly $\delta$ after the signals are launched from Port 1. Through the precisely timed switching, signals launched from Port 1 will be completely received by Port 2 without distortions. In the reverse paths, signals fed to Port 2 are rejected by Port 1's closed switches and received by Port 3 also through its timely turned-on switches. It is straightforward to notice that other ports are similar to Port 1 in terms of S-parameter performance due to the two-fold symmetry of the circuit. Consequently, non-reciprocal performance described by the following s-matrix can be achieved:

$$S = \begin{bmatrix} 0 & 0 & 0 & e^{-j\omega\delta} \\ e^{-j\omega\delta} & 0 & 0 & 0 \\ 0 & e^{-j\omega\delta} & 0 & 0 \\ 0 & 0 & e^{-j\omega\delta} & 0 \end{bmatrix} \quad (5)$$

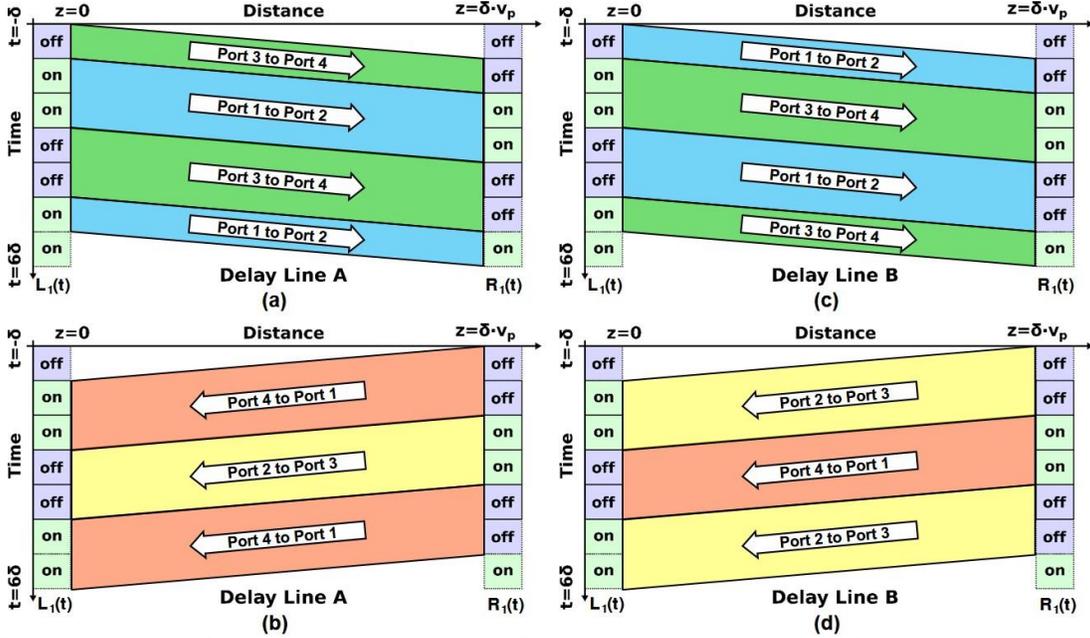

Fig. 3. Bounce diagram representation of 4-port circulation operation. Signal flow in delay line A is presented for (a) forward direction propagation (from left hand side to right hand side) and (b) backward direction propagation (from right hand side to left hand side). Signal flow in delay line B is presented for (c) forward direction propagation and (d) backward direction propagation. The columns on both sides of each diagram represent the states of switches $L_I(t)$ and $R_I(t)$ during this duration.

In essence, the time-reversal symmetry is broken through sequentially timing the switching from one side of the delay lines to the other side. It is also conceivable that such a framework can be programmable. Reassigning the control signals to ports would allow the signal to circulate in different directions. More detailed analysis on the programmability of such systems can be found in [25].

## B. Performance Degradation from Limited Delay Line Bandwidth

To understand the fundamental limit on the IL of such a system, we first study the transmitted signal in the time domain at different nodes of the system (Fig. 4) when only Port 1 is excited with an input signal of $f_A(t)$. Note that the analysis again assumes lossless and dispersionless delay lines and lossless switches with zero turn-on and off time.

Assuming the input signal to be time-harmonic at a frequency of $\omega_s$, the time-duplexed signals on the delay lines and combined signals can be seen at different nodes as:

$$f_B(t) = f_A(t) \cdot L_1(t) \quad (6)$$

$$f_C(t) = f_A(t) \cdot L_2(t) \quad (7)$$

$$f_D(t) = f_B(t-\delta) \cdot R_1(t) + f_C(t-\delta) \cdot R_2(t) = f_A(t-\delta) \quad (8)$$

$$f_E(t) = f_B(t-\delta) \cdot R_2(t) + f_C(t-\delta) \cdot R_1(t) = 0 \quad (9)$$

where A refers to the input node (Port 1), B and C refer to the nodes between the delay lines and the switches of Ports 1 and 3 on the left side, D refers to the output node, i.e. Port 2, and E refers to the isolated node, i.e. Port 4. These signals can also be represented in the frequency domain as:

$$L_1(\omega) = \sum a_n \cdot [\delta(\omega - n \cdot \omega_m)] \quad (10)$$

$$F_B(\omega) = F_A(\omega) * L_1(\omega) \quad (11)$$

$$F_C(\omega) = F_A(\omega) * L_2(\omega) = F_A(\omega) * [L_1(\omega) \cdot e^{-j\omega\delta}] \quad (12)$$

$$F_D(\omega) = F_A(\omega) \cdot e^{-j\omega\delta} \quad (13)$$

$$F_E(\omega) = 0 \quad (14)$$

where $\omega_m$ is the angular frequency of the switching/modulation signal ($\omega_m = \pi/2\delta$). $a_n$ is the Fourier series constant of the square wave. * is used to denote convolution. For a 50% duty cycle square wave, it can be expressed as:

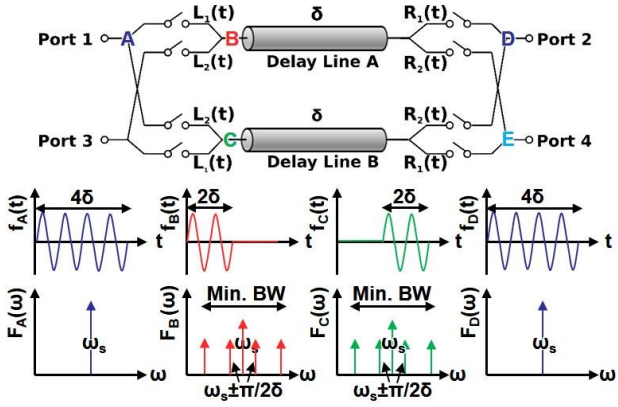

Fig. 4. Signal represented in both time domain and frequency domain at different nodes in the non-reciprocal network.

$$a_n = j/n\pi \cdot \sin(n\pi/2) = j/2 \cdot sinc(n/2) \qquad n < 0 \qquad (15)$$

$$a_n = 1/2 \qquad n = 0 \qquad (16)$$

$$a_n = -j/n\pi \cdot \sin(n\pi/2) = -j/2 \cdot sinc(n/2) \qquad n > 0 \qquad (17)$$

Eqs. 13 and 14 show that the transmitted signal is the input signal delayed by $\delta$, and no signal flows into the isolated port. It is noteworthy that, although the transmitted signal is the same time harmonic component as the input signal, the signals on both delay lines contain intra-modulated tones due to the time duplexing. The intra-modulated tones are defined as the mixed tones caused by the switching in the system.

As a result, the normalized bandwidth of the delay line (BW/$f_m$), or in other words the delay-bandwidth product (BW·$\delta$) of the delay line, plays an important role in setting the system performance. Two types of delay line responses can be commonly found, namely delay lines with band-pass (e.g. acoustic delay lines) and low-pass characteristics (e.g. EM planar transmission lines), which both will be used for our analysis. In our analysis, we also assume perfect filtering characteristics that include lossless and matched in pass band, infinite rejection in the stopband, infinitely steep transition between passband and stopband. The pass-bands and the spectrum of the time-duplexed terms are shown in Fig. 5 (a) – (b) for both cases respectively. Because of the limited BW of the delay lines, higher order intra-modulated tones are filtered by the delay lines, producing additional IL in the transmission paths of the circulator. For delay lines with a passband from $f_l$ to $f_u$, the IL can be calculated as:

$$IL = -10\log[\sum 2|a_n|^2 \cdot [H(f_s + nf_m - f_l) - H(f_s + nf_m - f_u) + H(f_s + nf_m + f_u) - H(f_s + nf_m + f_l)]] \qquad (18)$$

The IL is plotted in Fig. 5 (c) for the case of center frequency of the pass-band being much higher than the switching/modulation frequency. Predictably, steps are observed in Fig. 5(c) due to the discrete increase of number of the intra-modulated tones in the passband as the BW/$f_m$ expands.

The effect of a finite BW on IL is investigated for input signals at different frequencies. Generally, the circulator presents less IL to a signal in the center of the passband unless the BW is not enough to include the most adjacent pair of modulated tones. If a narrowband delay device (e.g. a two-port resonator) is used with a high modulation frequency, a minimum IL of 3 dB is expected due to the loss of energy in the modulated tones. A wider BW allows more tones to propagate from one end to the other end of the delay line. As a result, a larger delay-bandwidth product is desired in such a system.

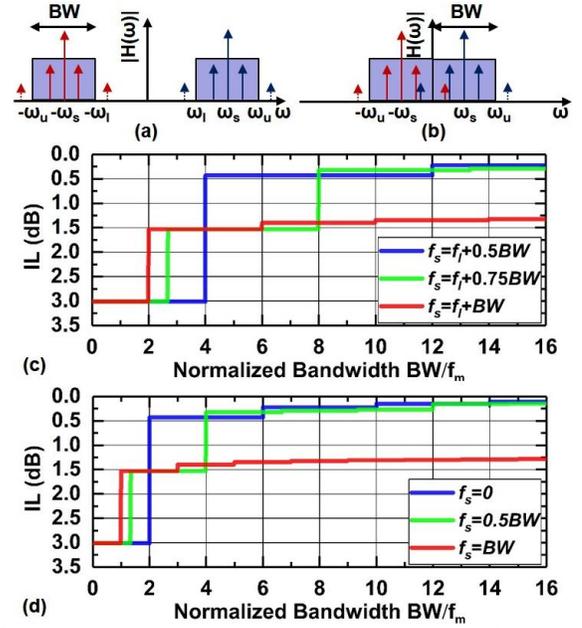

Fig. 5. Spectrum of the modulated signal on the delay lines with (a) band-pass characteristics and (b) low-pass characteristics. (c) Insertion loss as a function of delay-bandwidth product of the delay line (BW/$f_m$ or BW·4$\delta$). The effect of limited BW for input RF signals at different frequencies are plotted.

A similar analysis can be made for the low-pass case as the low pass case can be treated as a special case of the band-pass case by setting $f_l$ to 0. The IL can be calculated as:

$$IL = -10\log[\sum 2|a_n|^2 \cdot [-H(f_s + nf_m - f_u) + H(f_s + nf_m + f_u)]] \qquad (19)$$

The IL is plotted for different carrier frequencies in Fig. 5 (d), from which similar conclusions can be drawn. In general, a low-pass case allows more modulated tones due to the inclusion of the intra-modulated tones between $\omega_s$ and - $\omega_s$ in the pass-band. Obviously, a larger delay-bandwidth product is also helpful in such a system as well.

### C. Performance Degradation from Non-idealities

The previous subsection discusses the operation principles of a 4-port non-reciprocal network for the ideal case. In the analysis, the delay lines are wide band, lossless and matched to the system impedance with their delays perfectly synchronized to the switching cycle. The switches are assumed perfectly open in the off-state and short in the on-state without any rise time. The only loss in the transmission comes from the limited BW of the delay line. As a result, such an ideal 4-port non-reciprocal network provides infinite isolation. In implementation, the non-idealities will inevitably exist and degrade return loss, insertion loss and isolation of the circulator.

Different non-idealities can be sorted into two major categories. The first type includes the time-invariant losses and reflections caused by the delay line and switches. This covers the static insertion losses and reflection of the switches and the delay lines, which can be routinely analyzed using S-parameters of the non-ideal components. The performance degradation from static losses will be discussed in the Section III based on the selected components for implementation.

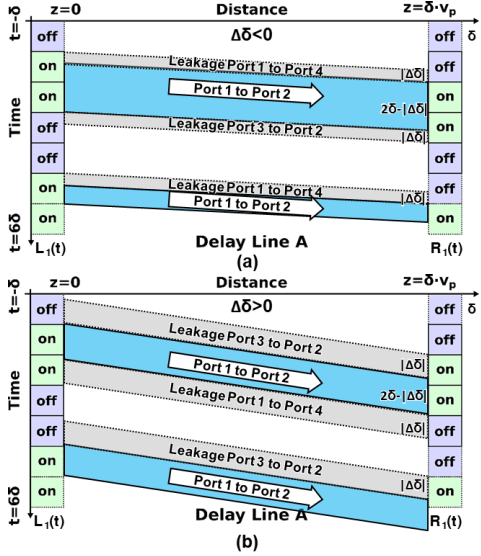

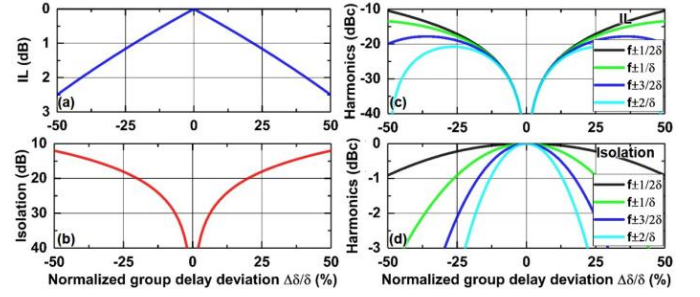

Fig. 6. Bounce diagram representation of the forward propagating wave (Port 1 to Port 2) in delay line A when delay deviation exists in both (a) $\Delta\delta < 0$ and (b) $\Delta\delta > 0$ cases. The state of switch $L_1(t)$ and $R_1(t)$ are plotted on the side. The signal leakage causes both a larger insertion loss and a worse isolation in both cases.

Fig. 7. (a) IL and (b) isolation of the system versus normalized delay deviation. Different orders of intra-modulated tones versus normalized delay deviation at the (c) through port and (d) isolation port.

The second type consists of losses and reflections associated with the time-varying nature of the system. In this subsection, we will only discuss the second type of non-idealities that are fundamentally a result of the off-synchronization between the switching and delays. Two types of off-synchronization can typically be found. The first type arises from either delay deviation or switching frequency deviation. Their effects are equivalent and can be analyzed simply with the delay deviation ($\Delta\delta$). The second type originates from a finite switching time ($t_s$), which leads to momentary loss and reflection of signals respectively during the turning-on or off periods.

For the first type, the delay deviation happens when the delay in the delay lines ($\delta + \Delta\delta$) and a quarter of the switch signal cycle ($\delta$) are different. In the analysis, $2\delta$ is assumed larger than $|\Delta\delta|$. The impact of delay deviation can also be described using the bounce diagram (Fig. 6) for two scenarios, when the delay in the delay line is shorter ($\Delta\delta < 0$) or longer ($\Delta\delta > 0$) than intended. For the shorter delay case ($\Delta\delta < 0$), the operation can be described in the time domain as:

$$T_1(t) = \sum [H(t/2\delta - |\Delta\delta|/2\delta - n) - H(t/2\delta - n - 1)] \quad (20)$$

$$T_2(t) = \sum [H(t/2\delta - n) - H(t/2\delta - |\Delta\delta|/2\delta - n)] \quad (21)$$

$$f_D(t) = f_A(t - \delta - |\Delta\delta|) \cdot T_1(t - \delta - |\Delta\delta|) \quad (22)$$

$$f_E(t) = f_A(t - \delta - |\Delta\delta|) \cdot T_2(t - \delta - |\Delta\delta|) \quad (23)$$

where $T_1(t)$ is a modified transmission waveform indicating the portion of the input signal transmitted to the output port. $T_2(t)$ is the complimentary of $T_1(t)$. Due to the delay deviation, a portion of the input signal is transmitted into the isolation port,

causing a higher IL and worse isolation, which can be presented in the frequency domain as:

$$T_1(\omega) = \sum t_{1n} \cdot [\delta(\omega - n \cdot 2\omega_m)] \cdot e^{-j\omega|\Delta\delta|} \quad (24)$$

$$T_2(\omega) = \sum t_{2n} \cdot [\delta(\omega - n \cdot 2\omega_m)] \cdot e^{-j\omega|\Delta\delta|} \quad (25)$$

$$F_D(\omega) = [F_A(\omega) \cdot e^{-j\omega(\delta + |\Delta\delta|)}] * [T_1(\omega) \cdot e^{-j\omega(\delta + |\Delta\delta|)}] \quad (26)$$

$$F_E(\omega) = [F_A(\omega) \cdot e^{-j\omega(\delta + |\Delta\delta|)}] * [T_2(\omega) \cdot e^{-j\omega(\delta + |\Delta\delta|)}] \quad (27)$$

$t_{1n}$ and $t_{2n}$ are the Fourier series constants of the modified transmission waveforms:

$$t_{1n} = j \cdot (1 - |\Delta\delta|/2\delta)/2 \cdot sinc[n(1 - |\Delta\delta|/2\delta)] \quad n < 0 \quad (28)$$

$$t_{1n} = (1 - |\Delta\delta|/2\delta)/2 \quad n = 0 \quad (29)$$

$$t_{1n} = -j \cdot (1 - |\Delta\delta|/2\delta)/2 \cdot sinc[n(1 - |\Delta\delta|/2\delta)] \quad n > 0 \quad (30)$$

$$t_{2n} = j \cdot (|\Delta\delta|/2\delta)/2 \cdot sinc[n(|\Delta\delta|/2\delta)] \quad n < 0 \quad (31)$$

$$t_{2n} = (|\Delta\delta|/2\delta)/2 \quad n = 0 \quad (32)$$

$$t_{2n} = -j \cdot (|\Delta\delta|/2\delta)/2 \cdot sinc[n(|\Delta\delta|/2\delta)] \quad n > 0 \quad (33)$$

For the longer delay case ($\Delta\delta > 0$), a similar analysis can be performed, leading to a similar conclusion. A larger delay deviation leads to more IL and worse isolation, which can be represented as:

$$IL = -20 \cdot log(1 - |\Delta\delta|/2\delta) \quad (34)$$

$$Isolation = -20 \cdot log(|\Delta\delta|/2\delta) \quad (35)$$

From Eq. 34 and 35, the effects of the delay deviation on the system IL and isolation are shown in Fig. 7 (a) − (b). It is interesting that the sum of IL and isolation is not unity. Part of the energy is converted into the modulated tones at the transmission port due to the delay deviation. The effect can be derived from Eq. 28-33 as:

$$Tran\_nth\_mod = 20 \cdot log(|sinc(n(1 - |\Delta\delta|/2\delta))|) \quad (36)$$

$$Iso\_nth\_mod = 20 \cdot log(|sinc(n(|\Delta\delta|/2\delta))|) \quad (37)$$

which is represented as the ratio of the the modulated tones to main tone (in dBc). The results are plotted in Fig. 7 (c) - (d). It is obvious that controlling control signal cycle to match the delay is important for the system. However, in a practical system with slight dispersive performance in the delay line passband, a certain degree of deviation is inevitable. To minimize the performance degradation, a longer delay in the delay line is necessary for larger tolerance.

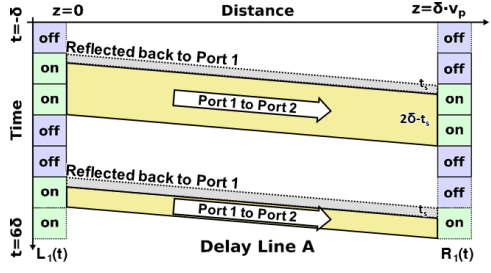

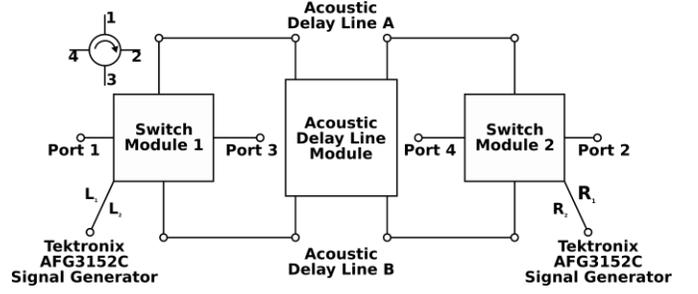

Fig. 8. Bounce diagram representation of the forward propagating wave (Port 1 to Port 2) in delay line A when switch rise time exists. The state of switch $L_1(t)$ and $R_1(t)$ are plotted on the side. The signal reflection in the switch causes both a larger insertion loss and a worse return loss.

Fig. 10. Block diagram of the constructed 4-port circulator, including two switching modules, and one delay line module that consists of impedance matching networks and unidirectional acoustic delay lines.

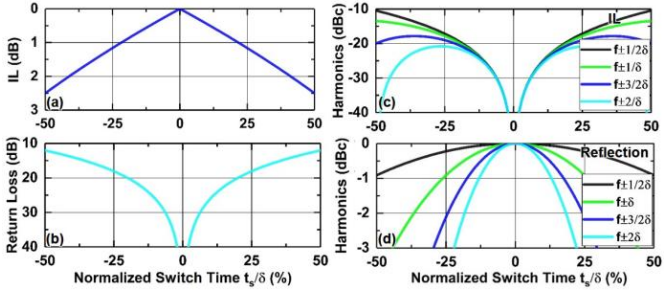

Fig. 9. (a) IL and (b) isolation of the system versus normalized switch time. Different orders of intra-modulated tones versus normalized switch time at the (c) through port and (d) isolation port.

The second off-synchronization arises from a non-zero switch time ($t_s$). Assuming a linearly varying impedance model for the switch with impedance changing from open to short during switching process [25], signal transmitted at the beginning of $t_s$ are reflected back to the input port (Fig. 8). In the analysis, $2\delta$ is assumed larger than $t_s$. The process can be described in the time domain as:

$$T_3(t) = \sum [H(t/2\delta - t_s/2\delta - n) - H(t/2\delta - n - 1)] \quad (38)$$

$$f_D(t) = f_A(t - \delta) \cdot T_3(t - \delta) \quad (39)$$

$$f_E(t) = 0 \quad (40)$$

where $T_3(t)$ is a modified transmission waveform indicating the portion of the input signal reflected. A portion of the input signal is reflected, causing more IL and worse isolation, which can be presented in the frequency domain as:

$$T_3(\omega) = \sum t_{3n} \cdot [\delta(\omega - n \cdot 2\omega_m)] \cdot e^{-j\omega t_s} \quad (41)$$

$$F_D(\omega) = [F_A(\omega) \cdot e^{-j\omega\delta}] * [T_1(\omega) \cdot e^{-j\omega\delta}] \quad (42)$$

where the Fourier series constants are:

$$t_{3n} = j \cdot (1 - t_s/2\delta)/2 \cdot sinc[n(1 - t_s/2\delta)] \quad n < 0 \quad (43)$$

$$t_{3n} = (1 - t_s/2\delta)/2 \quad n = 0 \quad (44)$$

$$t_{3n} = -j \cdot (1 - t_s/2\delta)/2 \cdot sinc[n(1 - t_s/2\delta)] \quad n > 0 \quad (45)$$

The analysis on the effects of the switch time on the system performance is similar to that for the group delay deviation. The results are plotted numerically in Fig. 9. The normalized switch time affects IL and RL, as well as the harmonics at the transmission port. To minimize the performance degradation, a longer delay is also preferable.

To summarize, the system performance is first analyzed in both the time and frequency domains for ideal cases. Next, the effects of limited bandwidth of the delay lines are analyzed. The analysis leads to the conclusion that wide bandwidth and long delays are essential in overcoming the fundamental limit on IL imposed by the filtering and dispersive effects of the delay lines. Afterward, the non-idealities related to the time-varying nature of our system have also been discussed. The effects of non-zero delay deviation and switch rise time normalized to the delay time indicate that a long delay provides higher tolerance to off-synchronization. Guided by the above analysis, we implement our non-reciprocal network by resorting to the best technological platforms that can provide a wideband, low IL, and a long delay, namely the LiNbO$_3$ acoustic delay lines. More details about the system implementation is shown in the next section.

## III. 4-PORT CIRCULATOR BASED ON SWITCHED ACOUSTIC DELAY LINES

### A. System Overview

Experimentally, we designed two standalone switch modules and one delay line module, and assembled them as the circulator (Fig. 10). The forward and backward directions of the circulator are labeled in the inset circulator symbol. 8 interfaces exist for such a system, including 4 signal ports and 4 control ports located on the switch modules. The switch modules are connected to each other through the delay line module with acoustic delay lines and matching networks. SubMiniature version A (SMA) connectors are used for both interface ports and interconnect ports. Similar to the topology in Fig. 2, Port 1 and Port 3 are on the same end of the delay lines, while Port 2 and Port 4 are on the other end. In operation, the control signals (synchronized and optimized for the delay) are fed into the switch modules while non-reciprocal performance is measured from Port 1 to Port 4.

In this section, we will first introduce and design of the switch module, and then present the standalone measurements for the switch module. Next, we will present the design and measurements of the fabricated delay lines and matching networks on the delay line module. Finally, we will show the synchronized control signals used for the system and introduce our circulator S-parameter simulation based on the key parameters measured from the subsystems.

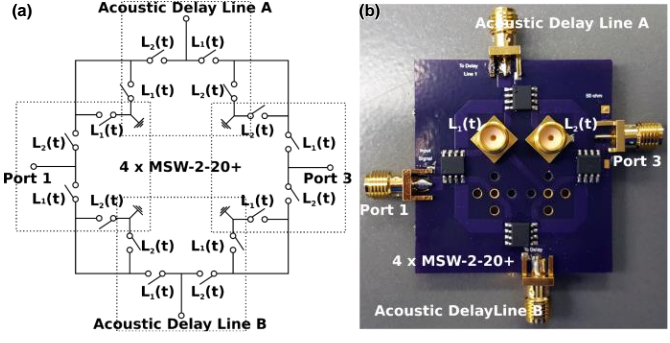

Fig. 11. (a) Schematic of the switching module. (b) Implemented switching module.

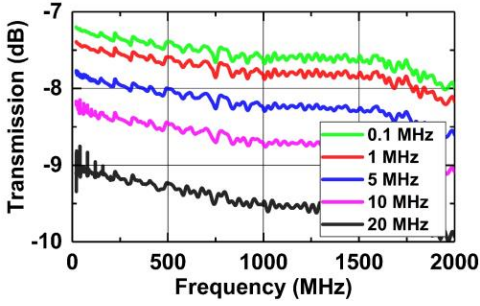

Fig. 12. Measured transmission Port 1 to Port acoustic delay line A with 50% duty cycle modulation signal in the switch module. The other two ports are terminated with matched load.

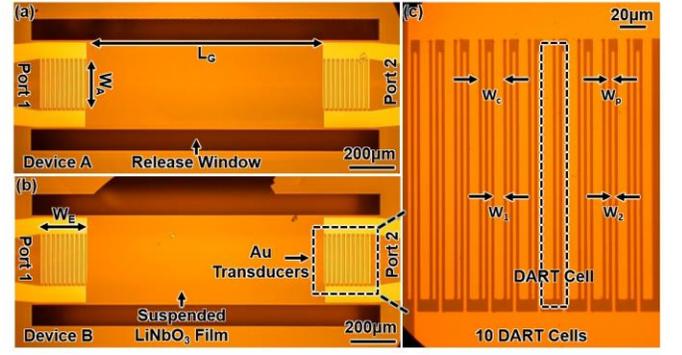

Fig. 13. (a) - (b) Optical microscope images of the fabricated acoustic delay lines with single-phase unidirectional transducers (SPUDT). A pair of distributed acoustic reflection transducers (DART) are arranged on both ends of the suspended LiNbO₃ thin film. A zoomed-in image is shown in (c).

TABLE I
DESIGN PARAMETERS OF LiNbO₃ SPUDT DELAY LINES

| Symbol | Parameter | Value |
|--------|-----------|-------|
| $L_T$ | Total length (mm) | 1.4 |
| $W_A$ | Aperture width (mm) | 0.2 |
| $W_E$ | TransducerW width (mm) | 0.2 |
| $L_G$ | Gap length (mm) | 1 |
| $W_c$ | Cell width (μm) | 20 |
| $W_1$ | Wider electrode width (μm) | 7.5 |
| $W_2$ | Narrower electrode width (μm) | 2.5 |
| $W_p$ | Pitch width (μm) | 2.5 |
| $T_{Au}$ | Gold thickness (nm) | 100 |
| $T_{LN}$ | LiNbO₃ thickness (nm) | 800 |
| $N$ | # of cells | 10 |

## B. Switch Module Design and Measurement

Two identical switch modules are incorporated in our system, but 180° rotated from each other. The schematic of the switch module is shown in shown in Fig. 11 (a). The board is composed of four signal ports (Port 1, Port 3, Acoustic delay line A and Acoustic delay line B) with interconnections between them through four single-pole, double-throw (SPST) switches (Mini-circuits MSW-2-20+). The switches are controlled by two control signals ($L_1(t)$ and $L_2(t)$, 180° phase difference). In operation, incident RF signals from Port 1 are sent to either Acoustic delay line 1 or Acoustic delay line 2, while signals from Port 3 are sent to the other acoustic delay line simultaneously. The extra switches to the ground enhance the isolation between Ports 1 and 3 due to additional reflection induced by shorting to ground.

The switch module is built on a 4.45-cm-wide, 1.6-mm thick FR-4 printed circuit board (PCB) as shown in Fig. 11 (b). Amphenol ACX1230-ND SMA connectors are used for the control signals and Amphenol ACX1652-ND SMA connectors are adopted for the RF signals. The SMA types are chosen for conveniently interconnecting the comprising parts of the system (Fig. 10). In the layout, the four RF ports are arranged symmetrically to ensure the signal paths between ports have the same electrical length and thus the same delay, which are essential for our non-reciprocal network. The ports for control signals are placed in the middle to minimize the phase difference between control ports and switches, and to reduce the crosstalk between the control signal paths and the RF signal paths. The dimensions of the microstrip transmission lines between the RF ports are optimized for lower insertion loss.

With the assembled switch module, the dynamic transmission loss between different ports are characterized with modulation control signals of different frequencies applied to the switch control ports. A Keysight N5249A PNA-X network analyzer is used to measure the IL between Port 1 and Port for Acoustic delay line A with the other two ports on the module terminated with matched loads (Mini-circuits ANNE-50+). A pair of control signals (0 to -3.3 V square wave, 180° phase difference) are provided as $L_1(t)$ and $L_2(t)$ from a Tektronics AFG3152C function generator. The measurement is shown in Fig. 12. An IL of 7.2 dB at 10 MHz is obtained with a 0.1 MHz modulation signal. According to Eq. 10 and Eq. 11, a minimum of 6 dB IL is expected for an ideal switch module measured in this dynamic manner due to the square wave modulation. The additional 1.2 dB loss is produced because of the IL of the switches (0.5 dB per switch from datasheet) and the interconnects (around 0.2 dB in total). A larger IL is obtained for higher frequency modulation signals due to a larger normalized switching time, as similarly shown in Fig. 9. Higher IL can also be observed for higher frequency RF signals (around 0.5 dB at 1.5 GHz) because of the loss in the FR-4 board.

## C. Design and Measurement of Delay Line Modules

The delay line module is built with the acoustic delay lines and the matching networks. The core of the delay line module is the in-house fabricated LiNbO₃ acoustic delay lines on a suspended X-cut single crystal LiNbO₃ thin film. The design is shown in Fig. 13. Each device is composed of input and output transducers with interdigitated electrodes that are on on top of suspended LiNbO$_3$ thin film, and separated by a delay distance. The device is oriented at -10° to +Y axis, at which the SH0 mode has the maximum $k_t^2$. In operation, the input transducers couple electrical energy into the acoustic domain and the output transducers convert the acoustic energy back to electrical signals, after a long group delay induced by the relatively low phase velocity of SH0 wave in LiNbO$_3$ (3200 m/s) [34]. As mentioned earlier, low IL in the delay lines is essential for meaningful implementation. Therefore, the conventional bi-directional transducer design in SAW with 6 dB minimum IL is not applicable [34]. In our work, we adopt single-phase unidirectional transducers (SPUDT) for eliminating the acoustic power transmitting away from the output transducers by phase-matching induced destructive interference. SPUDT transducers consist of a series of transducer cells with a width of (W$_c$). Each cell includes three electrodes (with the width of W$_c$/8, W$_c$/8, and 3W$_c$/8 respectively), which are placed $\lambda$/8 away from each other. The electrodes are alternatively connected to ground, signal and ground. The center frequency of the transducer is around the frequency at which the acoustic wavelength ($\lambda$) matches W$_c$. The BW of the transducer is collectively defined by the number of cells in a transducer and the required directivity. More details on the theory and analysis of the SPUDTs, and SPUDT-enabled delay lines based on LiNbO$_3$ thin films, are presented in [49].

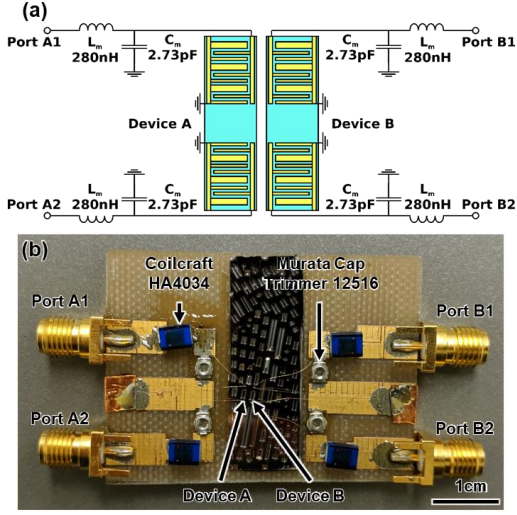

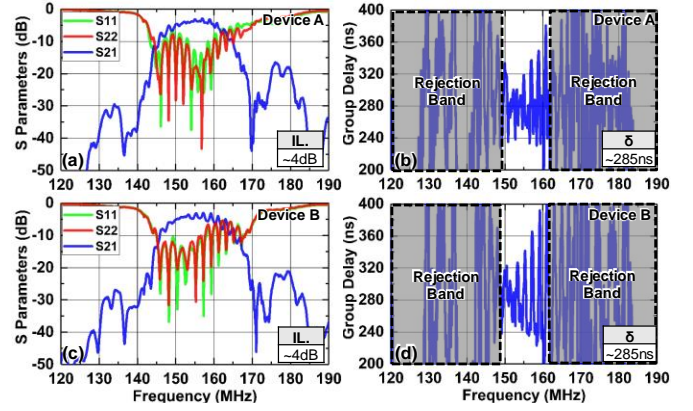

Fig. 15. Measured S-parameters and group delay of the matched (a) – (b) Delay line A and (c) – (d) Delay line B. An insertion loss around 4 dB and a group delay of 280 ns are measured.

Fig. 14. (a) Schematics of the 4-port delay line module consisting of two delay lines and four L-C impedance matching networks. (b) Delay lines assembled with matching networks on a FR-4 board.

TABLE II
LUMPED ELEMENTS FOR IMPEDANCE MATCHING

| Symbol | Parameter | Value |
|---|---|---|
| $L_m$ | Matching Inductance (nH) | 280 |
| $C_{m\_min}$ | Minimum Capacitance (pF) | 1.5 |
| $C_{m\_max}$ | Maximum Capacitance (pF) | 3 |
| $C_m$ | Matching Capacitance (pF) | 2.73 |

The devices were fabricated on a 1 cm by 3 cm LiNbO$_3$-on-Si sample with a process described in detail in [49]. The devices' center frequency is around 155 MHz (W$_c$ = 20 μm). The key physical parameters are presented in Table I. The fabricated devices show good uniformity across all transducer cells in the transducers with the designed electrode width [Fig. 13 (a) - (c)]. However, these devices cannot be directly connected with switches to form the non-reciprocal network due to their intrinsic complex input impedances and mismatch from 50 Ω. Therefore, impedance matching networks are designed with inductors and capacitors in a low-pass topology seen in Fig. 14 (a). As shown in Fig. 14 (b), a delay line module is subsequently implemented on a 1.5-mm-thick FR-4 board. An inductor and a tunable capacitor, with their parameters listed in Table II, are used for each matching network. The tunable capacitor is chosen for compensating the additional capacitance introduced by the wire-bonding. Symmetry in the design is also considered to ensure the same amount of group delays in both devices. The delay lines are measured with a Keysight N5249A PNA-X vector network analyzer (VNA). The performance of both acoustic delay lines is presented in Fig. 15 (a) – (d). The matched delay lines show an IL around 4 dB over a BW of 10 MHz and centered at 155 MHz. A group delay around 285 MHz is measured. The ripples in the group delay measurement are caused by the multi-reflections in the device, which are introduced by the finite directivity of the SPUDT transducers [49]. The group delay variation without considering multi-reflections have much fewer ripples [49]. It is important to note that such delay lines are the first of its kind and their performance is still far from ultimate. Nonetheless, their performance has already surpassed the state of the art SAW devices. With further improvement, sub-1 dB insertion loss is expected for the same delay.

### D. System Synchronization and Simulation

As seen in Fig. 2, four control signals with 90° phase difference are required in our system. In our implementation, these signals are generated by two synchronized two-channel Tektronics AFG3152C function generators. The generated signals are measured with an Agilent MSO7104B oscilloscope [Fig. 16 (a) – (d)], showing the required phase difference. To reduce the leakage from the control signals to the RF ports, Mini-circuits SLP-90 low-pass filters are used. The additional filters effectively remove the higher frequency overtones at the cost of a slightly longer rise-up time. The time domain measurement demonstrates a rise-up time of 6 ns (slightly larger than the switch rise-up time) [Fig. 16 (e)]. The spectrum of the control signal is measured by an Agilent E4445A spectrum analyzer [Fig. 16 (f)].

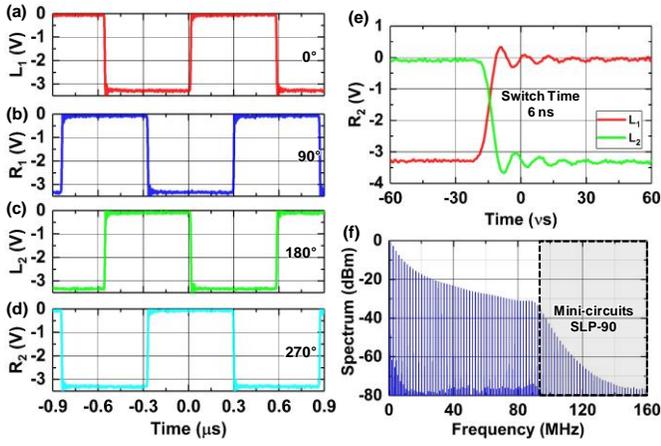

Fig. 16. (a) – (d) Synchronized control signals with 90° phase difference to each other measured with oscilloscope. (e) Zoomed-in time domain measurement of the control signals. A rise time of 6 ns is obtained. (f) Spectrum of the control signal. A low-pass filter is used to reduce the control signal leakage to the RF ports.

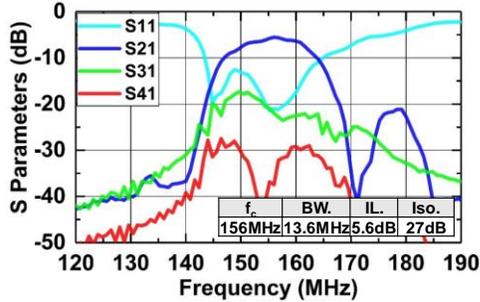

Fig. 17. Simulated S-parameters obtained from Advanced Design System. In the simulation, a switching time of 2 ns and on-state insertion loss of the switches are considered. Control signals are set to be 877.2 kHz (1.14 μs period).

A system level simulation is performed after characterizing the performance of individual modules. The simulation is based on a time-domain simulation in the Keysight Advanced Design System (ADS) [25]. In the simulation, a 6 ns switch rise time is assumed for each switch (Fig. 11), along with an on-state resistance of 3 Ω, and an off-state resistance of 60 kΩ. The delay line module is represented by the measured S-parameter (Fig. 15). Time domain simulations are performed with excitations of different frequencies from Port 1. Afterward, Fourier transforms are used to attain scattered power out of the other ports at the input frequency. An ideal circulator is used in the simulation to separate the incident wave and reflected wave at Port 1. As seen in Fig. 17, the simulation shows the anticipated non-reciprocal performance for our system. The performance is centered at 156 MHz with a BW of 13.6 MHz (FBW = 8.7%). An IL of 5.6 dB is measured with a great isolation of 27 dB. Compared to the frequency independent network [25], the BW of this system is limited by the BW of the miniature low-loss delay lines. The loss in the system is collectively caused by the time domain non-idealities (discussed in Section II) and conventional steady state losses, e.g. the switches, the delay line and the lumped elements in the matching networks. To break down the loss further, the 5.6 dB IL is caused by 4 dB loss in the delay line module (Fig. 15), and 1.6 dB loss caused by the two switch modules (Fig. 12). The 27 dB isolation can be further improved by reducing the port impedance mismatch at the interfaces between different modules and minimizing the delay deviation from the control signal.

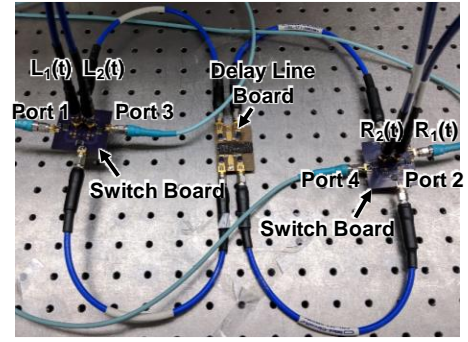

Fig. 18. Experiment setup of 4-port circulator, consisting of 2 switch boards and 1 delay line board. Interfaces are labeled on the figure.

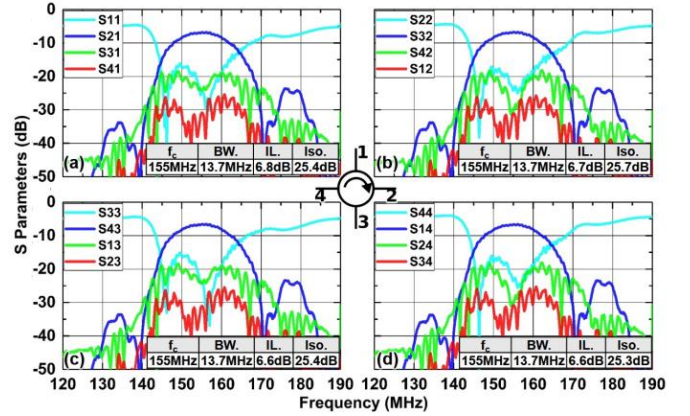

Fig. 19. Measured S-parameter performance of the 4-port circulator. Great performance symmetry is shown in the measurement. Minimum IL around 6.6 dB is measured at different ports. A bandwidth of 13.6 MHz (FBW=8.7%) has been obtained. Directivity larger than 19 dB is obtained between the forward and backward propagation path (e.g. between S12 and S21). Return loss is better than 15 dB at each port. The measurement is carried out at -10 dBm power level. Control signals are set to be 877.2 kHz (1.14 μs period).

TABLE III
INSERTION LOSS BREAKDOWN

| Source | Insertion Loss |
|---|---|
| Delay Line | 4 dB |
| Switch Module | 1.6 dB |
| Off-synchronization | 0.2 dB |
| Interconnections | 0.8 dB |
| Total | 6.6 dB |

## IV. MEASUREMENT AND DISCUSSIONS

### A. S-parameters and Group Delay

The whole system is assembled based on different modules discussed in Section III. SMA cables are adopted for interconnecting modules and feeding control and RF signals in switch modules (Fig. 11). The additional 0.8 dB degradation is caused by the loss in the SMA cables and connectors. Regarding the 25.4 dB isolation, the leaked control signal overtones and additional reflections cause the isolation degraded from the 27 dB isolation as predicted by the simulation (Fig. 17). Currently, the loss is limited by insufficient directionality in the SPUDT design and the lossy inductors in the impedance matching network, which will be significantly reduced by further optimization on acoustic delay lines with better designed cells and larger static capacitance for easier impedance matchings to 50 Ω systems.

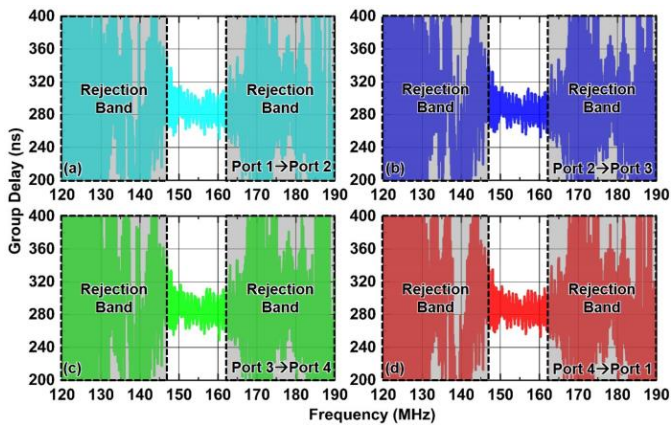

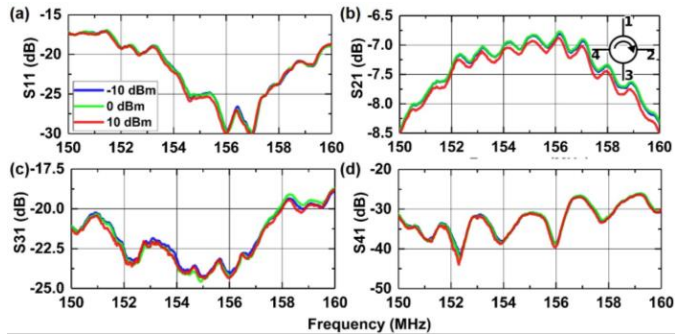

Fig. 22. Measured power handling of different S-parameters when Port 1 is excited at -10, 0, 10 dBm.

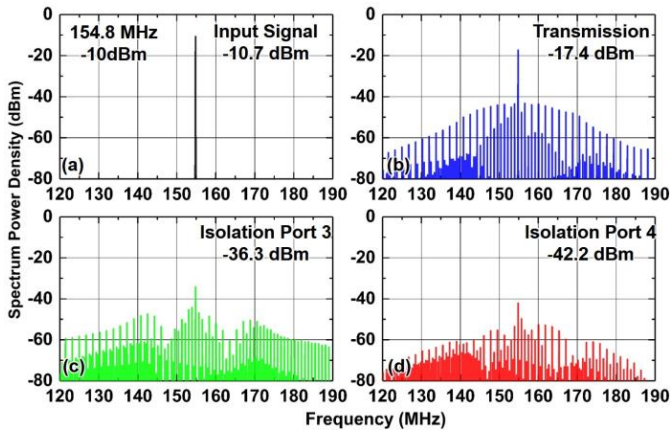

Fig. 20. Group delay measurement of different transmission paths. The propagation direction is shown in the right down corner of each subfigure. A delay of 285 ns is measured. The slight ripple in the delay is caused by the echoes in the delay lines caused by slight impedance mismatch.

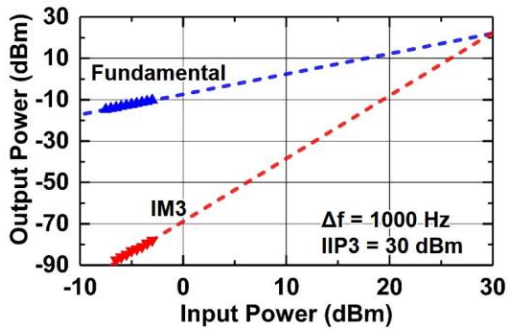

Fig. 23. Measured IIP3 of $S_{21}$ in the system. The frequency spacing is 1 kHz. An IIP3 of 30 dBm is obtained.

The group delay of the signals in the forward direction are presented in Fig. 20. Note that group delays around 285 ns are obtained with less pronounced ripples in comparison to stand-alone and static group delay measurements of the acoustic delay lines (Fig. 15). This is because the switching in circulator measurements actually mitigates the multi-reflections between ports and reduces group delay ripples.

### B. Intra-modulated Tones

Intra-modulated tones are measured with an Agilent N5181A signal generator and an Agilent E4445A spectrum analyzer. The -10 dBm, 154.8 MHz signal is input into port 1 [Fig. 21 (a)]. An IL of 0.7 dB is caused by the cables used for the measurement. The spectrum is measured when the other two ports are terminated with matched loads. A transmitted signal of -17.4 dBm is measured, showing an IL of 6.7 dB [Fig. 21 (b)]. The highest modulated harmonic is 26 dBc, which is caused by the delay deviation and slight off-sync in different switches. The isolated port (Port 4) shows a transmitted signal of -42.2 dBm [Fig. 21 (d)]. The highest modulated tone is the system (Fig. 18).

Fig. 21. Measured spectral content of (a) input signal, (b) transmitted signal at port 2, showing 6.5 dB insertion loss, (c) transmitted signal at port 3, indicating 25.4 dB isolation at port 3, and (d) transmitted signal at port 4, indicating 28.3 dB isolation at port 4. The intermodulation is caused by the non-ideal switching and multi-reflections in the spectrum.

First, S-parameters of the system are measured with a Keysight N5249A PNA-X network analyzer with four 877.2 kHz synchronized square waves applied to the switches. As shown in Fig. 16, the switching waveforms have a magnitude from 0 V to -3.3 V. The measured 4-port S-parameters (with the forward direction marked in the inset circulator symbol) show great performance symmetry among transmission and isolation between different ports (Fig. 19). The system also shows non-reciprocal performance for a BW of 13.6 MHz (8.9% FBW) at a center frequency of 155 MHz, which matches well with the system level simulation seen in Fig. 17. A minimum IL of 6.6 dB is obtained along with a minimum isolation of 25.4 MHz within the BW. As seen in Table III, the 6.6 dB IL can be further broken down for various causes. Around 4 dB arises from the acoustic delay line modules while 1.6 dB is from the switch modules. Both losses are considered in the simulation (Fig. 17). Around 0.2 dB is caused by the non-idealities in the switching synchronization and the slight phase offset caused by the unsymmetrical feed signal paths in the system.

TABLE IV
COMPARISON TO OTHER TIME-VARYING APPROACHES

| Reference | Center Frequency | Mod. Frequency | 3-dB BW* | Insertion loss | Isolation** | Intra-modulated Tones | IIP3 | Directivity | FBW | Symmetry | BW/Mod. Frequency |
|---|---|---|---|---|---|---|---|---|---|---|---|
| [13] | 130 MHz | 40 MHz | 5 MHz | 9 dB | 20 dB | N/A | N/A | 11 dB | 3.8% | Y | 0.125 |
| [20] | 750 MHz | 750 MHz | 32 MHz | 1.7 dB | 20 dB | N/A | 27.5 dBm | 18.3 dB | 4.3% | Y | 0.043 |
| [15] | 1167 MHz | 1 MHz | 2.2 MHz | 12 dB | 27 dB | N/A | 15 dB | 15 dB | 0.19% | Y | 2.2 |
| [16] | 145 MHz | 120 kHz | 0.2 MHz | 8 dB | 17 dB | N/A | N/A | 9 dB | 0.13% | Y | 1.67 |
| [17] | 2500 MHz | 3 MHz | 0.4 MHz | 11 dB | 40 dB | N/A | N/A | 29 dB | 0.016% | Y | 0.133 |
| [18] | 1000 MHz | 190 MHz | 2.4 MHz | 3.3 dB | 15 dB | 11.3 dBc | 33.7 dBm | 11.7 dB | 0.24% | Y | 0.012 |
| [19] | 1000 MHz | 100 MHz | 23 MHz | 2 dB | 23 dB | 29 dBc | 31 dBm | 21 dB | 2.3% | Y | 0.23 |
| [21] | 25 GHz | 8.33 GHz | 6 GHz | 3.2 dB | 15 dB | N/A | N/A | 11.7 dB | 24% | N | 0.72 |
| [24] | N/A | 6 MHz | 100 MHz | 5 dB | 40 dB | N/A | N/A | 35 dB | N/A | N | 16.7 |
| [25] | N/A | 23.8 MHz | 950 MHz | 5 dB | 30 dB | N/A | N/A | N/A | N/A | Y | 39.9 |
| **This work** | **155 MHz** | **877.2 kHz** | **13.7 MHz** | **6.6 dB** | **25.7 dB** | **26 dBc** | **30 dBm** | **19.1 dB** | **8.7%** | **Y** | **15.44** |

*Defined by IL. **Isolation within the BW.

around 8 dBc from the main tone.

### C. Power Handling and Non-linearity

The power handling capability of our system is collectively determined by the nonlinearities in the switches and the LiNbO₃ delay lines. The S-parameters are measured at different power levels (-10, 0, 10 dBm) with the Keysight N5249A PNA-X network analyzer. As seen in Fig. 22, the device shows good linearity for different S-parameters up to 10 dBm input power. Measurements at higher power levels were not performed due to the output power limit of the VNA.

Fig. 23 shows the IIP3 measurement. Two-tone signals are generated the Keysight N5249A PNA-X network analyzer and fed into Port 1. The output signal is obtained from Port 2 and measured with an Agilent E4445A spectrum analyzer. A frequency spacing of 1 kHz is set for the input signals. The measured IIP3 is 30 dBm, which is higher compared to other miniature MEMS piezoelectric resonators. The reason for the high IIP3 is that the delay lines are non-resonant devices and they do not confine energy as the MEMS piezoelectric resonators do [51], [52].

### D. Comparison to Other Time-varying Approaches

A full comparison of this work to the previously reported time-varying approaches is shown in Table IV. Compared to other works, our device demonstrates good directivity (isolation minus IL), good linearity, and excellent symmetry over a wide BW with a very low switching frequency of 877.2 kHz. A figure of merit, defined as the BW over modulation frequency, is also used to compare the obtained bandwidth for a given temporal effort (i.e. power in synthesizing modulation signals). We obtain a high FoM of 15.44, which was only previously attained from systems that are based on EM delay lines and have a much larger size. In addition to the high FoM, the intrinsic expandability and symmetry of our system imply that our approach would be beneficial to future high-performance MIMO systems.

Despite the IL and intra-modulations are still not as good as some of demonstrated approaches, the employment of acoustic delay lines with more integrated and faster switching will offer opportunities for significant future improvement. First, the delay line design can be optimized for better IL, and lower intra-modulations. The analysis in [49] demonstrates that the loss in the delay lines are from the transducers and insufficient transducer directionality, which can be drastically reduced with an optimized fabrication process (e.g. using aluminum electrodes with proper thickness). A reduction of IL in the delay lines from 4 dB to 2 dB is expected. Less group delay deviation is also expected from better directionality, which further leads to less pronounced modulated tones. Furthermore, due to the scalability of the delay lines, BW/$f_m$ can be improved with higher frequency delay lines. As long as the delay remains the same, the modulation frequency can remain the same and thus lead to no additional cost for synthesizing the control signals. Second, the switch module performance can be enhanced with lower loss and faster switches. Currently, 1.6 dB arises from the switch modules, which can be greatly reduced by better switch technologies with less static IL [53]. Better linearity can be also expected from a more advanced switching technology. Third, the 1 dB loss introduced by the SMA adapters and cables between modules can be eradicated with more integrated assembly between switches and delay lines. To sum up, a more compact circulator with sub-2 dB IL, better linearity, and less modulated tones is believe to be attainable with further optimizations.

## V. CONCLUSION

We presented the first non-reciprocal network based on switched acoustic delay lines. A quantitative analysis of the effect of non-idealities in the system on insertion loss, return loss, isolation, and intra-modulations is presented, concluding that long delays significant benefit the performance of such non-reciprocal systems. A 4-port circulator is then designed and implemented with two switch modules and the delay line module. The designs and the performance of different modules are individually analyzed, measured, and shown, before the circulator is assembled. The measurement shows a highly symmetric performance across the 4-ports with 18.8 dB non-reciprocal contrast between the IL (6.6 dB) and isolation (25.4 dB) over a FBW of 8.8% at a center frequency 155 MHz, all of which are accomplish with a record low switching frequency of 877.22 kHz. The system also shows 25.9 dB difference between the carrier and the intra-modulated tones, and IIP3 of 30 dBm. With the employment of faster switches, further optimizations on delay lines and synchronization, and more sophisticated integration, such circulators can potentially outperform ferrite-based devices in loss, bandwidth, and isolation while offering a more compact size and reconfigurable operation.

## ACKNOWLEDGMENTS

The authors would like to thank DARPA MTO NZERO and SPAR programs for funding support and Dr. Troy Olsson for helpful discussions.